\documentclass[runningheads,a4paper]{llncs}

\usepackage{amssymb}
\usepackage{graphicx}
\usepackage{epsfig}
\usepackage{amsmath}
\usepackage{authblk}
\usepackage{url}
\newcommand{\keywords}[1]{\par\addvspace\baselineskip
\noindent\keywordname\enspace\ignorespaces#1}

\begin{document}

\mainmatter  

\title{The Implications of Diverse Applications and Scalable Data Sets in Benchmarking Big Data Systems}
\titlerunning{Implications in Benchmarking Big Data Systems}
%
%
\author[1,2]{Zhen Jia}
\author[3]{Runlin Zhou}
\author[3]{Chunge Zhu}
\author[1,2]{Lei Wang}
\author[1,2]{Wanling Gao}
\author[1]{Yingjie Shi}
\author[1]{Jianfeng Zhan \thanks{The corresponding author is Jianfeng Zhan.}}
\author[1]{Lixin Zhang}
\affil[1]{State Key Laboratory Computer Architecture, Institute of Computing Technology, Chinese Academy of Sciences, China}
\affil[2]{University of Chinese Academy of Sciences, China}
\affil[3]{National Computer network Emergency Response Technical Team Coordination Center of China 
}
\authorrunning{Zhen Jia and etc.}
\institute{jiazhen@ncic.ac.cn, zhourunlin@cert.org.cn, jadove@163.com, wl@ncic.ac.cn, \{gaowanling, shiyingjie, zhanjianfeng, zhanglixin\}@ict.ac.cn}

%
%
%
\toctitle{The Implications of Diverse Applications and Scalable Data Sets in Benchmarking Big Data Systems}
\tocauthor{Zhen Jia, Runlin Zhou, Chunge Zhu, Lei Wang, Wanling Gao, Yingjie Shi,Jianfeng Zhan, and Lixin Zhang}
\maketitle
\begin{abstract}

Now we live in an era of big data, and big data applications are becoming more and more pervasive. How to  benchmark data  center computer systems running  big  data  applications (in short big data systems) is a hot  topic.
In this paper, we focus on measuring the performance impacts of diverse applications and scalable volumes of data sets on big data systems.
For four typical data analysis applications---an important class of big data applications, we find two major results through experiments: first, the data scale has a significant impact on the performance of big data systems, so we must provide scalable volumes of data sets in big data benchmarks. Second, for the four  applications, even all of  them use the simple algorithms, the performance trends are different with increasing data scales, and hence we must consider not only variety of data sets but also variety of applications in benchmarking big data systems.

\keywords{Big Data, Benchmarking, Scalable Data}
\end{abstract}

\section{Introduction}
In the past decades, in order to store big data and provide services,  more and more organizations around the world build data centers with scales varying from several nodes to hundred of thousands of nodes \cite{luo2012cloudrank}. Massive data are produced, stored, and analyzed in real time or off line. According to the annual survey of the global digital output by IDC, from 2005 to 2020, the digital data will grow by a factor of 300, from 130 exabytes to 40,000 exabytes.
The more data we produce, the more data center systems are deployed for running big data applications.

As researchers in both academia and industry pay great attention to innovative systems and architecture in big data systems \cite{barroso2009datacenter} \cite{zhan2012high} \cite{ferdman2011clearing} \cite{HVCBench} \cite{lotfi2012scale} \cite{setting} \cite{jianli},
the pressure to evaluate and
compare performance and price of these systems rises \cite{ghazal2013bigbench} \cite{baru2013benchmarking}.
Benchmarks provide fair basis for comparison among different big data systems. Besides, benchmarks represent typical needs of system support from big data applications. Together with workload characterization of typical big data applications, benchmarking results can thus enable active improvements of big data systems.

In a tutorial given at HPCA 2013 \cite{HPCA_2013}, we stated our position on big data benchmarking: we should take an incremental approach in stead of a top-down approach because of the following four reasons: first, there are many classes of big data applications, and there is a lack of a scientific classification of different classes of big data applications.  Second, even for data center workloads, there are many important application domains, e.g., search engines, social networks, though they are mature, customers, vendors, or researchers from academia or different domains of industry  do not know enough to make a big data benchmark suite because of the confidential issues \cite{bigdatachen}. Third, the value of big data drives the emergence of innovative application domains, which are far from our reach.  Fourth, the complexity,
diversity, scale, workload churns, and rapid evolution of big data systems indicate that both customers and vendors often have incorrect or
outdated assumptions about workload behaviors \cite{bigdatachen}.  Recently, big data benchmarking communities make a first but important step, and Ghazal et al.  present BigBench, an end-to-end big
data benchmark proposal \cite{ghazal2013bigbench}, whose underlying business model
of BigBench is a product retailer. Although we have some insights of the big data applications \cite{jiacharacterization} \cite{chenzheng} \cite{zhan2012high},  
considering the challenges mentioned above, there is a long way to go.

Workload, application and data are all important for characterizing big data systems \cite{sang2012precise}.
In this paper, we focus on data analysis workloads---an important class of big data application, and investigate the performance impacts of diverse applications and scalable volumes of data set in benchmarking big data systems. We choose four typical data analysis applications from a benchmark suite for  big data systems \cite{gaoisca}, and use different input data sets,
the scale of which ranges from Mega Byte to Tera Byte, to drive those applications. As Rajaraman explained \cite{rajaraman2008more}, for big data applications, inferior algorithms beat better, sophisticated algorithms because of the computing overhead.  The four applications we chose indeed use simple algorithms, whose computation complexities slightly vary from $O(n)$ to  $O(n \times log_2n)$. We use a user-perceived performance metric---data processed per second to depict the system processing capability.

Through experiments, we learnt that:
\begin{itemize}
	\item For the four representative big data applications, data scale has a significant impact on the performance of big data systems, so we must provide scalable volume of data sets in big data benchmarks.
	\item For the four representative big data applications, the performance trends are different with increasing data scales, and hence we must consider not only the variety of data sets but also the variety of applications when benchmarking big data systems. This also implies that there is no one-fit-all application.
\end{itemize}

The remainder of the paper is organized as follows. Section \ref{expirement} shows the workloads and evaluation methodology.
Section \ref{result} reports the experiment results and Section \ref{further} gives our analysis. Section \ref{lesson} discusses the
implications of our observations in benchmarking big data systems.  
Section \ref{conclusion} draws conclusions and mentions the future work.

\section{Evaluation Methodology}\label{expirement}
\subsection{Workloads}

We choose four representative Hadoop applications from BigDataBench\cite{gaoisca} including
\emph{Sort}, \emph{Word Count}, \emph{Grep} and \emph{Naive Bayes}.

\emph{Sort} is a representative I/O-intensive application, which simply uses the MapReduce framework to sort records within a directory.
\emph{Word Count} is a representative CPU-intensive application, which reads text files and counts how often the words occur.
\emph{Grep} is frequently used in data mining algorithm, and it extracts matching strings from text files. 
\emph{Naive Bayes} is a simple probabilistic classifier which applies the Bayes'
theorem with strong (naive) independence assumptions.

In this paper, these four applications we chosen all have relatively low computational complexity.
This is because that ``More data usually beats better algorithms" \cite{rajaraman2008more}.
Table \ref{app} shows some details of the four applications.
\begin{table}
\caption{Details of  Different Algorithms}\label{app}
\center
\begin{tabular}{|c|c|c|}
  \hline
  Application & Time Complexity & Characteristics  \\ \hline
  Sort & $O(n \times log_2n)$ & Integer comparison  \\ \hline
  WordCount & O(n) & Integer comparison and calculation  \\ \hline
  Grep & O(n) & String comparison  \\ \hline
  Naive Bayes & $O(m \times n)$ & Floating-point computation  \\
  \hline
\end{tabular}
\end{table}

\subsection{Performance Metric} \label{metric}
We adopt a user-perceived performance metric - data processed
per second to reflect the system's data processing capability.
For each application, the metric of data processed per second is defined as the input data size divided by the application running time. For example, the running time of \emph{Sort} with 100 GB input data set is 2487 seconds, and then the data processed per second of \emph{Sort} is 41.6 MB/s.
For \emph{Sort}, this metric means the application can sort 41.6 Mega Byte data per second.

In order to explain the trend of each application's processing capability, we also collect several
micro-architectural and operating system level metrics.
We get the micro-architectural data by using hardware performance counters.
We use Perf---a profiling tool for Linux 2.6+ based systems \cite{perf}, to
drive the hardware performance counters collecting micro-architectural events.
In addition, we access the \emph{proc} file system to collect OS-level
performance data, such as the I/O wait time.
We collect all the four slave nodes data, and report the mean value.

\subsection{Summary of Hadoop Job Execution \cite{apachehadoop}}
The four applications are all based on Hadoop.
Hadoop is a framework that allows for the distributed processing of large data sets using the Map/Reduce model \cite{apachehadoop}.
A MapRedcue job consists of a map function and a reduce function, and Hadoop breaks each job into tasks.
Each map task processes one input data block (typically 64 MB) and produces intermediate results. Reduce tasks deal with the
list of intermediate data through the reduce functions and produce the jobs' final output \cite{apachehadoop}.
Job scheduling is performed by the unique master node of Hadoop, and there are also many slave nodes which own a fixed number
of map slots and reduce slots to run tasks. The master assigns tasks of the job in response to heartbeats sent by slaves,
which report the number of free map and reduce slots on the slave \cite{zaharia2010delay}.
In our experiments, we submit the Hadoop jobs one by one and use the default FIFO scheduler policy. So the tasks  of each job will be queued in the master node and be executed in FIFO orders too.

\subsection{Experiment Platforms}
We use a 5-node cluster to run those applications. Each node has two Xeon E5645 processors equipped with 16 GB memory and 8 TB disk. For the 5-node cluster, we deploy a Hadoop environment on it (1 master and 4 slavers).
The details of configuration parameters of each node are listed in Table \ref{hwconfigeration}.
\begin{table}
\caption{Details of Configurations}\label{hwconfigeration}
\center
\begin{tabular}{|c|c|}
  \hline
  CPU Type & Intel \textregistered Xeon E5645\\ \hline
  \# Cores & 6 cores@2.4G \\ \hline
  \# threads& 12 threads \\ \hline
	\#Sockets & 2 \\ \hline
  \hline
  L1 DCache & 32KB, 8-way associative, 64 byte/line \\ \hline
  L1 ICache & 32KB, 4-way associative, 64 byte/line \\ \hline
  L2 Cache & 256 KB, 8-way associative, 64 byte/line \\ \hline
  L3 Cache &  12 MB, 16-way associative, 64 byte/line \\ \hline
  Memory & 32 GB , DDR3 \\  \hline
	Network & 1 Gb ethernet link\\ \hline
\end{tabular}
\end{table}

The operating system is Centos 5.5 with Linux kernel 2.6.34.
The Hadoop version is 1.0.2, and the java version is JDK 1.6.
For each slave node, we assign 18 map slots and 18 reduce slots 
with 512 MB Java heap for each slot.  
For other Hadoop configurations, we use the default ones.

\section{Evaluation Results and Analysis}\label{result}
\subsection{Data Scale}
For those four applications, we use different input data sets to drive those applications. For \emph{Sort}, the scale of the input data sets ranges from 200 MB to 100 GB. For \emph{Word Count} and \emph{Grep}, the scale of the input data sets ranges from 200 MB  to 1 TB, respectively.  For \emph{Naive Bayes}, the scale of input data sets ranges from 160 MB to 300 GB.
In order to eliminate the experiment deviations, each experiment is performed at least two times. We report the mean values across several times experiments.

\subsection{Experiments Observations} \label{ob}

\begin{figure}
\centering
\includegraphics[scale=0.5]{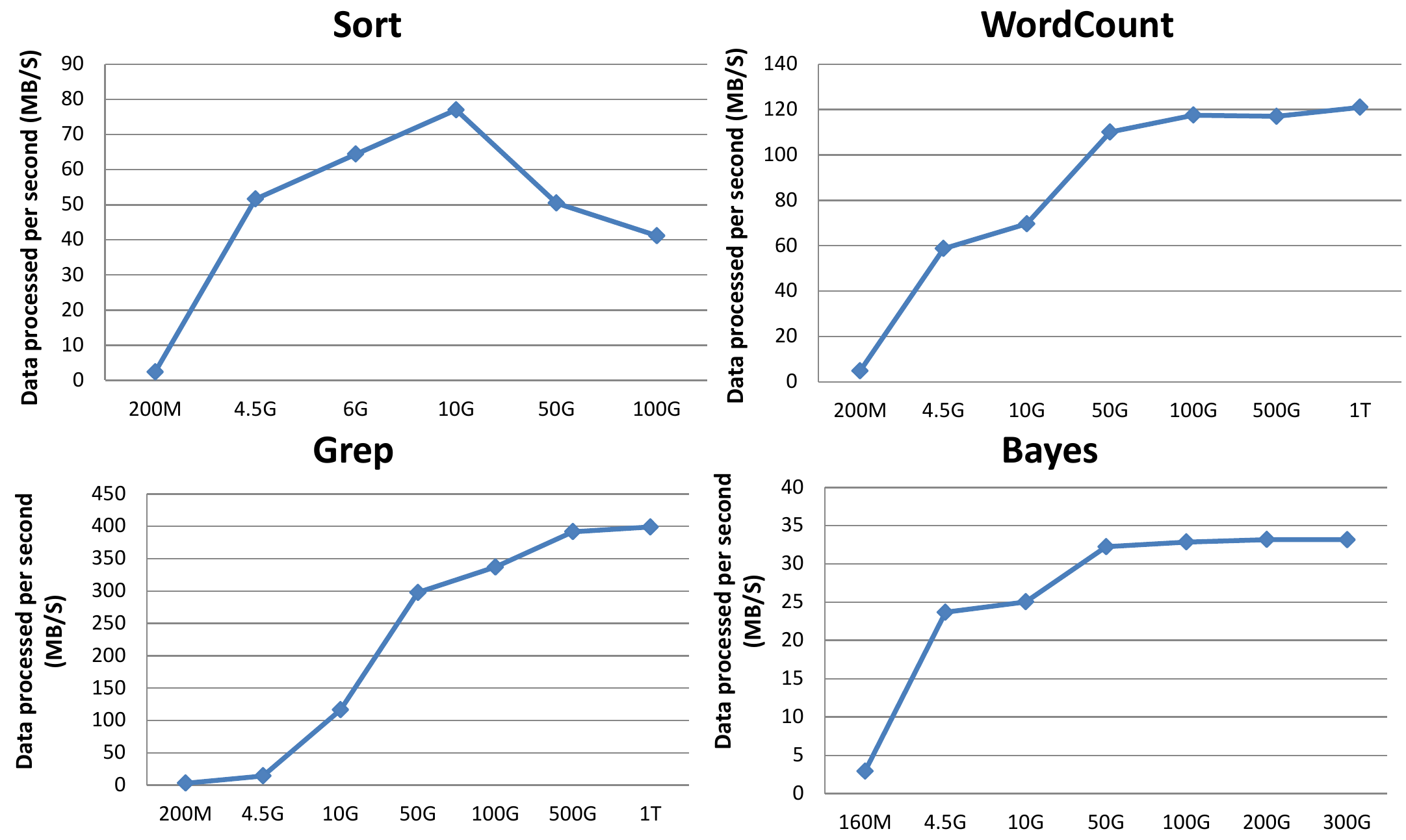}
\caption{The System's Processing Capability}\label{compare}
\end{figure}

Figure \ref{compare} shows the system's data processing capability, which is the performance metric
defined in section \ref{metric}. 
We can find that
the system has significantly different data processing capabilities when running different applications with different scale of data sets.
For example, the system processing capability running \emph{Grep} is more than 3 times than that of running \emph{WordCount} when they both process 1 TB data set.
Meanwhile, the performance metrics of big data applications are sensitive to the data scales.
Even for the same application, the processing capability is significantly varied from different scales of input data sets.
For example, running \emph{Grep}, the performance of the system  is 3.077 MB per second  when the data scale is 200 MB, while the processing capability  is up to 398.7 MB per second with 1 TB data input.
The details of our findings from those experiments are described as follows.

First, different applications have different processing capabilities. We can find that the maximum processing capability is 336 MB/s (Grep) and the minimum processing capability is 33 MB/s (Naive Bayes)
when the data scale is 100 GB ,respectively. This is because  that \emph{Naive Bayes} classifies records based on a
probability model. It needs to calculate posterior probability for each record. So it is the most
time-consuming one in the four applications, and has the lowest processing capability. 
While \emph{Grep} is much simpler than the other three applications. It only finds the matched strings in each record, so it has the highest processing capability.
In Section \ref{further}, we provide more performance data to explain this observation.
Those different processing capabilities of different applications imply that varieties of workloads must be considered in big data benchmarking,
since a certain application can not represent the behaviors of all workloads in big data field.
A benchmark suite composed of diverse workloads is needed.

Second, the same application has different processing capability with different data scales. For all the four applications, we can find that
there is a stage where the processing capability increases with increasing of data scale.
This can be seen as a process of stressing the system step by step, which leads the system to a 
state of resource being fully used, and hence a peak system processing capability will appear.
The reason why applications' processing capabilities increase with increasing of data scale is that the computing resources are not fully used when the data set is small, especially when the data size is less than the 4.5 GB.
The basic data block size for each Hadoop map task is 64 MB in our experiments \cite{white2012hadoop}, 
so on our Hadoop cluster, the minimum data size driving all map slots to run tasks concurrently is 4.5 GB
($64~MB \times 18~map~tasks \times 4~slaves$). When the input data set is too small (less than 4.5 GB),
the Hadoop will just allocate some of map slots to complete the job.
This situation causes only some of slaves busy and others less busy or even idle. So when the data set is less than 4.5 GB, applications show low processing capabilities. When all the map slots are used, the processing capabilities increase.
After testing with 4.5 GB data set, we use larger data sets to stress the system further.
We can find there is a turning point, of which processing capability curves stop increasing: 10 GB for \emph{sort}, 500 GB for \emph{Grep}, 100 GB for \emph{WordCount} and 50 GB for \emph{Naive Bayes}, respectively.
There may be some fluctuations, which are within the range of allowable deviation.
This phenomenon can be caused by many reasons, such as the different computational complexities, diverse resource requirements, and diverse system's bottlenecks, which will be further explained in Section \ref{further}.
The highest points in the figure mean the maximum processing capability in our experiments.
The corresponding abscissa value is the data set which can drive applications to reach the
maximum processing capability. The phenomenon implies that we should tune the scalable volume of input data set to achieve the peak performance.
What we must point out is that the data set size, which drives the system to reach the maximum processing capability, is an approximation for we do not enumerate all the data set size in our experiments.
Take \emph{Sort} for an example. In our experiment environment, the highest point is at 10 GB point. The input data set size, which can drive the application to reach the maximum processing capability, is about 10 GB.
However, the 10 GB is an approximation, for we do not know whether a 9 GB data set or an 11 GB data set can achieve better processing capability.
For the other three applications, their processing capability curves tend to smooth along with the data scale increasing. It implies that the maximum processing capabilities of them are near to the smooth points of them.

\section{Further Analysis} \label{further}

This section will analyze the causes of phenomena in Figure \ref{compare}.
We will find the main factors, which cause the processing capability varying with data scale.
First, we will report the cluster's resource requirements with data scale increasing, and
then investigate whether the computational complexity theory can explain the processing capability trend.
At last, we will explain some interesting phenomena. 

\subsection{Resource Requirements}
As mentioned in section \ref{ob}, increasing the input data size is a process of stressing the system and
using more resources step by step.
Resource consumption characteristics have great influence on the application performance \cite{zhancost} \cite{wang2012cloud},
so we would like to investigate the resource requirements and resource utilization for each application.
The operation system level metrics can reflect applications' requirements directly since the operating system is the one that manages hardware resources and provides services for applications running upon it.

For an application can be decoupled into data movement and calculating, the operating system level metrics we choose are I/O wait percentage and CPU utilization, which can reflect the data movement and calculating.
We get those metrics from the \emph{proc} file system as mentioned in Section \ref{metric}.
We collect the system time, user time, irq time, softirq time and nice time, and sum those time up as the CPU used time.
The CPU utilization is defined as the CPU used time divided by all CPU time.
The I/O wait percentage is defined as the I/O wait time, which can also get from the \emph{proc} file system, divided by all CPU time.

I/O wait time means the time spent by CPU waiting for I/O operations to complete.
A high percentage of I/O wait time means that the application has I/O operations frequently, which further indicates that the application is an I/O intensive workload.
For system, high I/O wait implies that I/O operations may be the system's bottleneck.
The CPU utilization reflects how much time the CPU is used to do calculation instead of waiting for I/O or idle.

\begin{figure}
\centering
\includegraphics[scale=0.6]{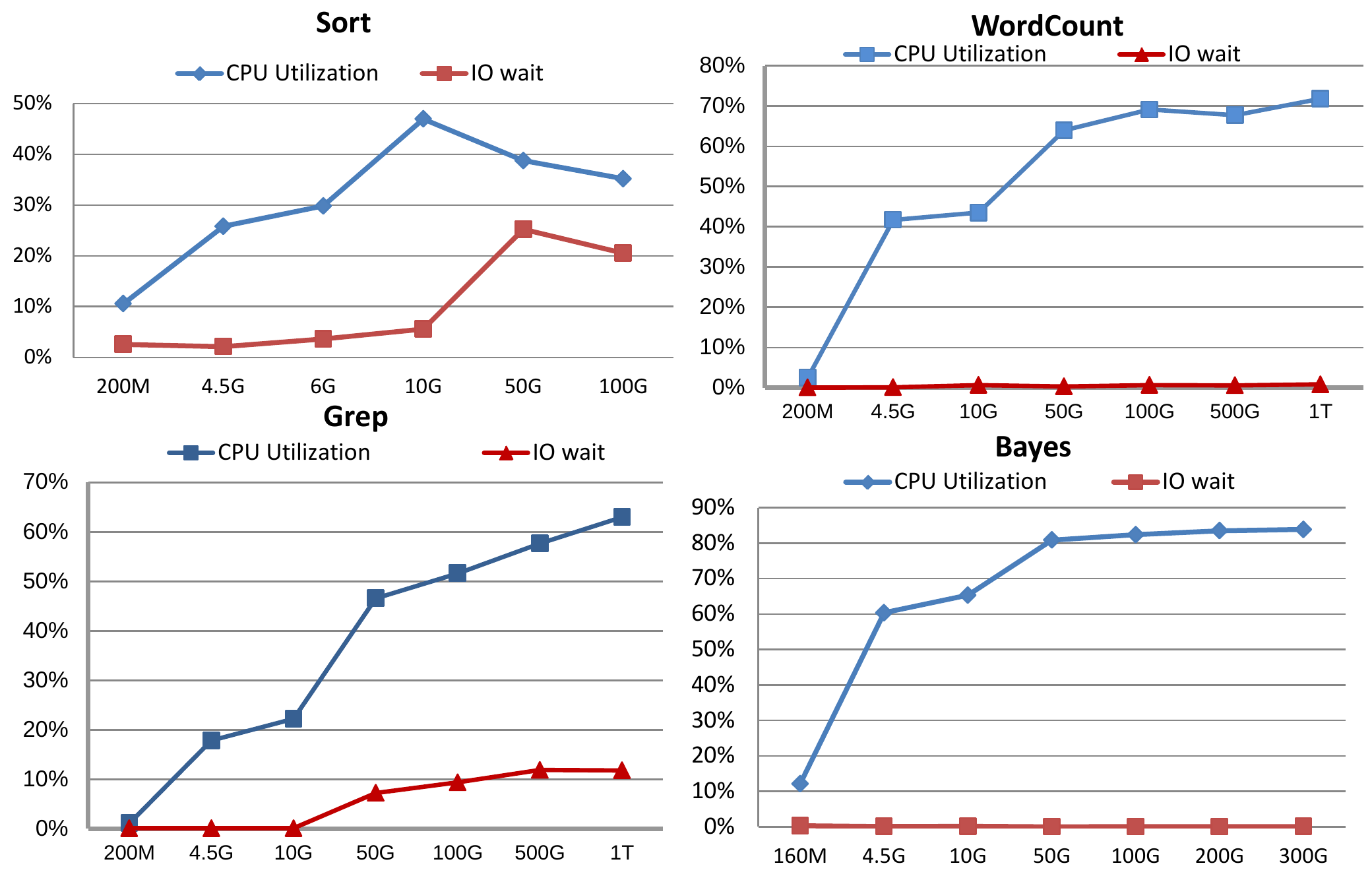}
\caption{The CPU Utilization and I/O Wait Percentage of Each Workloads}\label{analysis}
\end{figure}
Figure \ref{analysis} shows the CPU utilization and I/O wait time percentage of each workload.
For \emph{Sort}, when the data size is less than 10 GB, the data processing capability increases with the data scale increasing, and the CPU utilization goes up for it uses the Hadoop slots more efficiently.
When the data size is larger than 10 GB, the processing capability decreases with data scale
increasing. From Figure \ref{analysis}, we can find that the system's I/O wait time increases intensively whereas the CPU utilization decreases when data scale is greater than 10 GB.
This phenomenon means that system is waiting for the data coming and further decreases the processing capability.
The last point of \emph{sort} application in
Figure \ref{analysis} seems strange. At 100 GB point the  CPU usage decreases and the I/O wait time
decreases at the same time, which seems unreasonable. This phenomenon is caused by the unbalanced
I/O wait time \footnote{We run the \emph{Sort} 100 GB data set several times. Each time the experiment has the similar phenomenon.}.
The data we showed in Figure \ref{analysis} is the average value of the four slaves.
For the four slaves, we find that the maximum I/O wait time percentage is 31.5\% and minimal
I/O wait percentage is 17.6\% with the average 25.3\% in the face of 50 GB data. Whereas the 100GB point's maximum I/O wait
percentage is 36.6\% and minimal I/O wait time percentage is 10.9\% with average 20.5\%.
The variance of running 50 GB data set is 27 whereas the variance
is 94 for running 100 GB data set, which indicates that running 100 GB data set makes the system more unbalanced.

So here we can find that for the I/O intensive application -- \emph{Sort},
the processing capability trend is mostly impacted by the
I/O operations. The large percentage of I/O wait time elongates the \emph{Sort}'s execution time and further reduces the processing capability. The I/O operation becomes a bottleneck for \emph{Sort} application.

Different from \emph{Sort}, the other three applications (\emph{WordCount}, \emph{Grep} and \emph{Naive Bayes}), are not I/O-intensive applications, and they do not have an obvious bottleneck.
So the processing capability is mostly decided by CPU
utilization.
When the system's resource is fully used, the processing capability is unchanged.

\subsection{What about Computational Complexity Theory?}

The computational complexity theory is used to identify the inherent difficulty of solving a problem，
and it is also interested in the time consuming with an increase in the input size, which matches our
scenario.
The time required to solve a problem with certain scale is commonly expressed using big O notation, which is called
time complexity.
Such as we showed in Table \ref{app},
the time complexity of \emph{Sort} algorithm is $O(n \times log_2n)$.
The time complexity of  \emph{Grep} and \emph{Wordcount} is $O(n)$, and
the time complexity of \emph{Naive Bayes} is $O(m \times n)$, where \emph{m} is the length of
dictionary. The \emph{m} is a constant, so the complexity can also be seen as $O(n)$.
For the complexity, researchers actually use the RAM (Random Access Machine) \cite{cook1973time} model to measure it for the Turning Machine method is incredibly tedious \cite{complexity}.
In order to calculate the time complexity, the researchers need to analyze the source code of the application, and find the operations in the execution path.
In RAM model, 
each simple operation takes exactly one time step.
Each memory access takes exactly one time step.
Under the RAM model, the running time of an application is measured by counting up the number of time steps 
taking on a given input data set \cite{skiena1998algorithm}.

The RAM time-complexity is calculated by counting basic arithmetic operations in source code. 
And the compiler will compile the source code to instructions according to the processor's
ISA (Instruction Set Architecture).
The number of instructions executed can reflect how much work the processor need to do.
So we collect the number of instructions executed for each workload.
We calculate the instructions
executed per Mega Byte data processing by using formula \ref{eq2}. The reason why we use this metric is that the three out of our
four applications own time complexity of $O(n)$.
When an algorithm's time complexity is $O(n)$, the number
of instructions executed should increase in the proportion of increasing data scale.
That is to say, if we double the data scale, the number of executed instructions should also be doubled.
So the instructions executed per Mega Byte data should be unchanged, when the application faces different data sizes.

\begin{equation} \label{eq2}
\frac {\sum{^{slave 4}_{slave 1}} ~Instruction~ executed}{Input ~Data ~Size~(in ~Mega~ Byte~)}
\end{equation}

\begin{figure}
\centering
\includegraphics[scale=0.5]{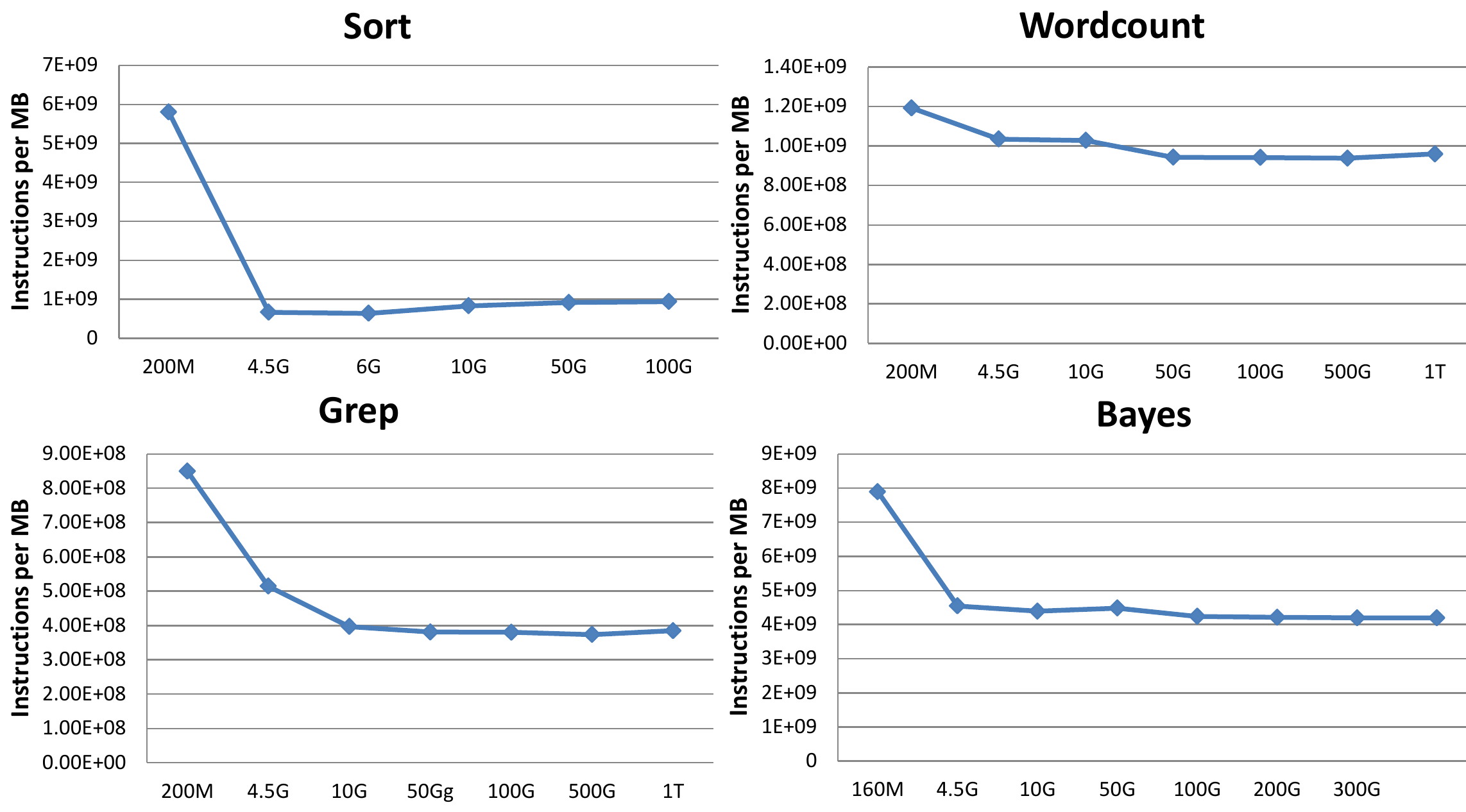}
\caption{Instructions Executed per Mega Byte Data Processing of Each Application}\label{ins}
\end{figure}

Figure \ref{ins} shows the number of instructions executed for processing each Mega Byte data.
We can find that when the input data set is small, such as 200 MB, 4.5 GB and etc., the number of
instructions executed for processing each Mega Byte data is more than that of larger input data sets. 
This is because the four applications all use Hadoop framework.
The framework will introduce extra instructions, such as the demon process \emph{TaskTracker}
and \emph{DataNode} will execute many instructions.
The extra instructions will affect the metric, instructions executed per Mega Byte data,
especially when the input data set is small.
When the data set is less than 4.5 GB,
even though some slaves do not run application tasks, the Hadoop framework instructions will also be executed and
counted, such as instructions  executed by \emph{DataNode} process. And hence the percentage of Hadoop framework instructions of small data set driven workload is much larger than that of  workload driven by large data set.
So an application driven by a small data set will execute more instructions per Mega Byte data than the large data sets.
This can explain why the curves in Figure \ref{ins} decrease sharply from the smallest data set to
4.5 GB data set.
When enlarging the input data set, the application will execute
more application instructions and further amortize the extra instructions introduced by Hadoop framework.
For the four applications, the stationary points are different. 
For \emph{Grep} the stationary point is 10 GB, and for the other three applications the stationary
points can be seen as 4.5 GB. Although there are some fluctuations, they are within the range of
allowable deviations.
The different stationary points are caused by different logic the four applications own.
The \emph{Naive Bayes} is the most complex one as mentioned in Section \ref{ob}. 
So it needs the maximum number of instructions to process each Mega Byte data among the four applications (about $4.2 \times 10^9$ instructions executed for processing each Mega Byte data).
The large amount of instructions needed for processing each Mega Byte data make it easy to amortize
the instructions introduced by Hadoop framework.
Whereas the \emph{Grep} is the simplest one among the four applications. It needs the minimum number of
instructions to process each Mega Byte data
(about $3.8 \times 10^8$ instructions executed Mega Byte data).
So it needs more application instructions to amortize the Hadoop framework introduced instructions.
That's why the stationary point for \emph{Grep} is 10 GB, while the points for other three are 4.5 GB.
After the stationary points in Figure \ref{ins},
we find the instructions executed for each Mega Byte data fit the complexity theory.
The instructions executed for processing each Mega Byte data remain unchanged, when the application faces different scale data.

In Figure \ref{ins}, the \emph{Sort} has the same trend with other three applications, even though its
time complexity is $O(n\times log_2n)$.
This can be explained by using the following interpretation.
Let us assume that we enlarge the data set $x$ times for \emph{Sort}. The time complexity will be $x \times n log_{2}(x \times n)$.
The complexity increases $x+log_nx$ times. which can be explained by
equation \ref{equ1}.
The $n$ in equation \ref{equ1} is \emph{Sort} application's record number. For \emph{Sort}, each record size is about 10 KB on average. For a 10 GB input data set the record number
is about 1 million, whereas the $x$ is not a big number.
The $x$ is 5 when the data set increases from 10 GB to 50 GB.
The $x \times log_{n}x$ will be a very small number.
So the time complexity increases can be seen as $x$,
which implies the total number of executed instructions increases at liner rate with data scale increasing.
So the instructions executed for processing each Mega Byte data nearly remain the same, which is consist with Figure
\ref{ins}.
We can conclude that the four applications' instructions executed situation meets the complexity
theory.
\begin{equation} \label{equ1}
\begin{aligned}
&~~~~\frac{x \times n \times log_2(x\times n)}{n\times log_2n} \\
&= \frac{x \times (log_2x+log_2n)}{log_2n}\\
&=x + x\bullet \frac{log_2x}{log_2n}\\
&=x+x \bullet log_{n}x
\end{aligned}
\end{equation}

\subsection{Additional Interesting Phenomena}
Besides the above discussions, we also find some interesting phenomena, we will show the phenomena and 
explanations in the rest of this section.

\textbf{Phenomenon 1:}
The \emph{sort}'s processing capability trend decreases sharply when the data scale is larger than 10 GB in Figure \ref{compare}.

\textbf{Explanations:}
According to those applications' time complexity,  the \emph{Sort}'s processing capability should remain unchanged or decrease slightly after the resource is fully used.
For the data processing capability can be evaluated as
$n/(n \times log_2n)=1/log_2n$. The processing capability will
decrease at the speed $x \bullet log_nx$, which is a very small number just as explained above.
But we find that \emph{Sort} application's processing capability decreased sharply when data set is larger than 10 GB in Figure \ref{compare}. 
The processing capability
decrease between 10 GB data set and 50 GB data set reaches 53\% (the process capability for 10 GB data
and 50 GB data are 77.01 MB/s and 50.45 MB/s respectively).
Whereas the instructions needed for processing each Mega Bytes data nearly remain unchanged (Figure \ref{ins})
when the data set is larger than 10 GB for \emph{Sort}.
This phenomenon is caused by the RAM (Random Access Machine) model,
which is used in calculation the time complexity.
As mentioned above, the RAM  model assumes that
each simple operation takes exactly one time step.
Each memory access takes exactly one time step, and we have as much memory as we need.
The RAM model is too simple, which covers up many real situations,
such as division two
numbers takes more time than adding two numbers in most cases, memory access times differ greatly
depending on whether data sit in cache or on the disk and etc \cite{skiena1998algorithm}.
So the RAM model can not depict the time consumed accurately,
especially the long latency memory access. If the data is not in main memory, it will take a long time waiting  for data coming and the I/O wait time is increased.
That is to say, the long latency memory access, will cause the CPU waiting for the data coming.
During this time, the instructions, which are waiting for the operand, will not be executed until the data come.
The instruction is delayed and further the corresponding operation will need more time to complete.
For \emph{Sort} application, the long I/O wait time elongates the instruction execution time and further decrease the processing capability.
From Figure \ref{analysis}, we can find that, the I/O wait time percentage increase with the
data increasing. The long I/O wait time extends the instructions execution time and makes the processing capability trend deviate from the complexity trend.
Just as Larry Carter found that
the performance looks much closer to $O(n^{5})$ instead of $O(n^{3})$ when doing  matrix multiply on
IBM RS/6000 \cite{matrix}.

For the other three applications (\emph{WordCount}, \emph{Grep}, \emph{Naive Bayes}), they do not have an obvious bottleneck with the data scale increasing.
Although  operations and memory access do not take the same time step,
the average time of processing each record tends to convergence when data volume is large enough for each application.
That's why those three applications' processing capability trends meet the time complexity.

\textbf{Phenomenon 2:}
In Figure \ref{compare}, different applications have different processing capabilities even
though they process the same amount of data and have the same time complexity.

\textbf{Explanations:}
The complexity theory is used to direct algorithm design, 
instead of evaluating 
the processing capability among different kinds of algorithms.
The value of time complexity (big \emph{O} notation expressed) is an estimated value.
The big \emph{O} notation expressed time complexity is said to be described asymptotically, i.e., as the input size goes to infinity. It only includes the highest order term and excludes coefficients and lower order terms.
The complexity calculated as function of the size of the input. It can give a trend of time consuming with the scale increasing when face certain problem in theory. Even though the trend may deviate from
real situation, it can be used to direct algorithm design.
For example,
when facing the same problem, such as classification, an $O(n^2)$ algorithm is worse than an $O(n)$
algorithm e.g. \emph{Naive Bayes}, in most  instances.
However, for different problems, the operations mix can be different, and the number of basic operations needed for processing each unit of data is also different. For instance, when processing the same amount of data, the instructions needed for \emph{Grep} and \emph{WordCount} are totally different in Figure \ref{ins}.
So the time complexity can not be used to evaluate different kinds of algorithms.

Actually the processing capabilities are mainly decided by the instructions executed per Mega Byte data and the systems bottleneck after the system resource is fully used.
For instances, when process 100 GB data,
\emph{Grep} needs 0.39 Tera instructions whereas \emph{Naive Bayes} needs 428 Tera instructions even though they all have the computational complexity of O(n).
So the \emph{Grep} has better processing capability than \emph{Naive Bayes}.
The \emph{Sort}'s processing capability is 77.01 MB/s when it processes 10 GB data, whereas it is 50.45 MB/s when facing 50 GB data. This is because that the percentage of I/O wait time is enlarged and becomes a bottleneck.

\textbf{Phenomenon 3:}
Different applications' highest processing capability appears at different data scales in Figure \ref{compare}. 

\textbf{Explanations:}
Different applications have different resource requirements.
\emph{Naive Bayes} needs more CPU resources than \emph{WordCount}.
When they both process 10 GB data set, \emph{WordCount}'s CPU utilization is 43.94\% whereas the
\emph{Naive Bayes}'s is 65.23\% (in Figure \ref{analysis}).
The more CPU resources needed by \emph{Naive Bayes} drive it to reach
the highest point faster.
This phenomenon can explain why the definitions
of "large" and "small" depend on the specific applications \cite{Chen:EECS-2012-174}.

\section{Lessons Learnt From the Experiments} \label{lesson}

Through the above experiments, we learnt several lessons in benchmarking big data systems.

\subsection{Consider the Scalable Volumes of Data Inputs in Big Data Benchmarking}

The data scale has a significant impact on the performance evaluation of big data systems.
Even for the same application, the processing capability of the big data system in terms of data processed per second  varies significantly with increasing data scales.
For example, running \emph{Grep}, the processing capability of the system  is 3.077 MB per second  when the data scale is 200 MB, while the processing capability  is up to 398.7 MB per second with 1 TB data input.
If we want to benchmark a big data system, the system should be fully used, only in
this way can the system show peak performance. Big data is needed for stressing test big data systems. 
In addition, larger data set can reduce the impacts from framework. As mentioned in Section \ref{further}, large data set can amortize the framework introduced instructions and further decrease the framework's impacts.

From \emph{Sort} application, we can also conclude that big data requires big data system.
When we enlarge the \emph{Sort}'s input data set, the processing capability decreases sharply for
the large proportion of I/O wait time. It is too inefficient to process big data by using a small scale
system. 
The phenomenon can also explain why more data usually beats better algorithms \cite{rajaraman2008more} in some degree. The big data can stress the bottleneck of the system such as I/O operations
for \emph{Sort}, so the algorithms designed for processing big data should
pay more attention to avoid system's bottleneck instead of reducing the time complexity only.

In order to benchmark big data systems, we must tune the volumes of data inputs so as to get the peak performance of the system and reduce the impacts of framework, and hence scalable volumes of data input must be provided in big data benchmarks.

\subsection{Consider Diversities of Workloads in Big Data Benchmarking}

Also, we find that, running different applications results in varied performance number even they use the same scale of data input.
For example,  the processing capability of running \emph{Grep}  is more than 3 times that of running \emph{WordCount} when they process 1 TB input data set.

As Baru et al. \cite{setting} mentioned, big data issues impinge upon a wide range of applications, covering from scientific to commercial applications. Different applications have different processing
capabilities. It is difficult to single out one application to represent all.
So when we evaluate big data systems, we must consider not only variety of data sets \cite{ghazal2013bigbench}, but also variety of workloads.
Different workloads can also reduce the impact of a specific application.
Our previous work shows that customizable workloads suite is preferred to meet users' requirements \cite{luo2012cloudrank}.

\subsection{The Limitation of the Sort Benchmark}
Lastly, the state-of-practice methods for big data systems evaluation, such as
MinuteSort\cite{apacible2012minutesort}, JouleSort, GraySort and TeraByte Sort \cite{sort}, have their limitations, since most of  them own a fixed scale of data input.

For example, TeraByteSort reports the performance with a 1 TB data input, which only reflects its sort performance with a 1TB data. But we do not know its performance when the data scale increases up to 10 TB or 1 PB. At the same time, we do not know whether
the 1 TB data can drive the system to achieve the maximum processing capability.
Another example is MinuteSort.
If the MinuteSort's result of a big data system is 100 GB, which reflects that it sorts specific 100 GB data in one minute. But we do not know the processing capability in the face of 1 TB data.

Moreover, the sort benchmarks only consider one algorithm and fail to cover the diversity of workloads in big data fields.

\section{Conclusion and Future Work}\label{conclusion}
In this paper, we paid attention to an important class of big data applications---data analysis workloads. Through the experiments we find that first, the data scale has a significant impact on the performance of big data systems, so we must provide scalable volumes of data sets in big data benchmarks so as to achieve peak performance for big data systems with different scales. Second, for the data analysis workloads, even all of  them use the simple algorithms, the performance trends are different with increasing data scales, and hence we must consider not only variety of data sets but also variety of applications in benchmarking big data systems.

For data analysis workloads,
we adopt an incremental approach to build benchmark suite.
Now we have investigated application domains, singled out the most important applications and
released a first version benchmark suite \cite{gaoisca} on our web page (http://prof.ict.ac.cn/BigDataBench).
In the near future, 
we will continue to add more representative benchmarks to this suite.
Especially, we will also develop data
generation tools, which can generate scalable volumes of data sets for big data benchmarks.

\section*{Acknowledgment}
We are very grateful to anonymous reviewers. This work is supported by the Chinese 973 project (Grant No.2011CB302502), the Hi-Tech Research and Development (863) Program of China (Grant No. 2011AA01A203,
 2013AA01A213), the NSFC project (Grant No.60933003, 61202075), the BNSF project (Grant No.4133081) and the 242 project (Grant No.2012A95). 

\bibliographystyle{abbrv}
\bibliography{tex}
\end{document}